\documentclass{article}
\usepackage{spconf,amsmath,graphicx,multirow,multicol,subfig}
\usepackage{booktabs}
\usepackage{amsfonts}

\title{LightGrad: Lightweight Diffusion Probabilistic Model for Text-to-Speech}
\name{
Jie Chen$^{1,\dagger}$, Xingchen Song$^{2,3}$, Zhendong Peng$^{2,3}$, Binbin Zhang$^{2,3}$, Fuping Pan$^2$, Zhiyong Wu$^{1,4,*}$
\thanks{$\dagger$ Work done during an internship at Horizon Robotics.}
\thanks{* Corresponding author.} 
}
\address{
    $^1$ Shenzhen International Graduate School, Tsinghua University, Shenzhen, China\\
    $^2$ Horizon Robotics, Beijing, China \quad $^3$ WeNet Open Source Community\\
    $^4$ The Chinese University of Hong Kong, Hong Kong SAR, China\\
    \small{
        chenjie20$@$mails.tsinghua.edu.cn,
        binbin.zhang$@$horizon.ai,
        zywu$@$sz.tsinghua.edu.cn%,
    }
}
\begin{document}
\ninept
\maketitle

\begin{abstract}
% 1. TTS发展，催生了大量TTS应用，为满足这些需求，大量TTS模型部署在云端
% 2. 云端模型不能满足隐私保护和低延时需求，需要能够部署在端侧的模型
% 3. diffusion模型不断发展，它训练稳定，而且参数更高效
% 4. 但是应用diffusion模型到端侧面临两大问题：
%    （1）模型参数量不够lightweight
%    （2）去噪步数过多，时延很高
% 5. 提出我们的工作：基于GradTTS的Lightweight模型，降低参数量，推理步数，同时使用流式推理加快推理速度
% 6. 说一下最后的效果如何
% 7. demo与模型代码的位置
Recent advances in neural text-to-speech (TTS) models bring thousands of TTS applications into daily life, where models are deployed in cloud to provide services for customs.
Among these models are diffusion probabilistic models (DPMs), which can be stably trained and are more parameter-efficient compared with other generative models.
As transmitting data between customs and the cloud introduces high latency and the risk of exposing private data, deploying TTS models on edge devices is preferred.
When implementing DPMs onto edge devices, there are two practical problems.
First, current DPMs are not lightweight enough for resource-constrained devices.
Second, DPMs require many denoising steps in inference, which increases latency.
In this work, we present LightGrad, a lightweight DPM for TTS.
LightGrad is equipped with a lightweight U-Net diffusion decoder and a training-free fast sampling technique, reducing both model parameters and inference latency.
Streaming inference is also implemented in LightGrad to reduce latency further.
Compared with Grad-TTS, LightGrad achieves 62.2\% reduction in paramters, 65.7\% reduction in latency, while preserving comparable speech quality on both Chinese Mandarin and English in 4 denoising steps\footnote{Demos and code are available at: https://thuhcsi.github.io/LightGrad/}.

\end{abstract}


\section{Introduction}

% 1. 现有tts系统的综述：非轻量化模型
% 2. 讲一下diffusion模型的工作，如GradTTS，WaveGrad，DiffWave
% 3. 轻量化模型的意义，轻量化模型的工作的介绍
% 4. 说一下diffusion模型没有轻量化的语音合成模型，讲一下diffusion模型在应用到端侧的时候面临的两个问题：模型参数量，去噪推理步数造成的高时延
% 5. 说一下我们proposed的模型：
% （1）对参数进行了轻量化，14M->5M出结果
% （2）应用了一种不同的方法来加速推理，并且提供了实验对比其效果
% （3）加入流式推理优化

Text-to-speech (TTS) converts text into intelligible and natural speech audios. 
With the development of deep learning, neural network based TTS models thrived and greatly simplified the complex TTS pipelines.
Although autoregressive neural TTS models, such as Tacotron \cite{tacotron} and Transformer TTS \cite{transformertts} have shown superior performance, they suffer from slow inference speed and a lack of robustness.
Later, non-autoregressive neural TTS models were introduced to overcome disadvantages of autoregressive ones.
FastSpeech \cite{fastspeech} improved inference speed and robustness by estimating token lengths and generating speech frames in parallel.
FastSpeech2 \cite{fastspeech2} was proposed with a simplified training pipeline and a novel variance adaptor better solving the one-to-many mapping problem in TTS.
To eliminate the necessity of an external aligner and further simplify the training procedure of non-autoregressive TTS models, Glow-TTS \cite{glowtts}, a flow-based generative model equipped with monotonic alignment search (MAS) was introduced.

To meet the growing demand for TTS in daily life such as virtual assistant, screen reader and navigation, neural TTS models, such as those mentioned above are widely deployed in cloud to provide services.
However, this paradigm is now being challenged by privacy protection and low-latency requirement.
First, transmitting data to the cloud may expose private information.
Second, sending data between cloud and customs makes TTS applications sensitive to network condition and increases user-perceived latency.
Thus, lightweight neural TTS models which can be deployed on edge devices are both desirable and necessary.
Previous works in lightweight TTS model have explored optimizing the architecture of FastSpeech \cite{luo21,li21,Xiao22}, using convolution network as main building blocks \cite{Kang21,Jan20}, adopting semi-autoregressive mode when synthesizing \cite{wang21}, designing lightweight on-device autoregressive TTS model \cite{devicetts} and combining variational generator with flow to enjoy benefits of both \cite{Ren21}.

%zywu-1006 已经解决
%要类似第一段的逻辑，这里要说明DPM解决的是什么问题？为什么要提出DPM？它有什么好处。
%对于SDE也是类似的。
Recently, a new class of generative models called diffusion probabilistic models (DPMs) emerges \cite{ddpm1,ddpm2}, which uses a parameterized Markov chain trained to gradually convert a simple distribution into complex data distribution.
Compared with other generative models, such as generative adversarial networks (GANs) and flow-based generative models, DPMs can be stably optimized and are more parameter-efficient.
Previous works have demonstrated that DPMs can produce promising results in image generation \cite{ddpm3}, neural vocoder \cite{diffwave,wavegrad}, TTS \cite{gradtts,difftts}, singing voice synthesis \cite{diffsinger}, and voice conversion \cite{diffsvc,gradvc}.
To apply new sampling methods and extend capabilities of DPMs, \cite{sde} proposed to use stochastic differential equation (SDE) to describe the forward process and backward processe of DPMs.
Grad-TTS \cite{gradtts} shows that transforming data distribution into $\mathcal{N}(\mu,\mathit{\Sigma})$ instead of $\mathcal{N}(0,\mathit{I})$ in forward process of DPMs can improve the overall quality of synthesized speech and reduce computation needed to generate satisfying results in inference.

However, lightweight DPM for TTS has not yet been explored.
Deploying DPMs on edge devices has two practical problems.
%zywu-1006 已经解决
%这里的逻辑没问题，但是用词、写法不太准确。
%computation resources和#parameters能对应么？
%many steps的hight inf. latency与computational complexity好像也不能对应？
First, present DPMs for TTS are not lightweight enough to be deployed on edge devices.
Second, DPMs usually need many denoising steps to get satisfying results in inference, which can introduce high inference latency.
Thus, a successful implementation of DPM on edge devices has to reduce both model parameters and inference latency.

% 这一段的内容会根据实验结果做调整
In this work, we present LightGrad, a lightweight DPM for TTS which has much smaller model size and faster inference speed.
To reduce model parameters, we propose a lightweight U-Net decoder, where regular convolution networks in diffusion decoder are substituted with depthwise separable convolutions \cite{depthwise}.
To accelerate the inference procedure, we adopt a training-free fast sampling technique for DPMs \cite{dpmsolver}.
In addition to a reduction in denoising steps, streaming inference is implemented in our model to reduce inference latency further \cite{Ellinas20}.
Compared with Grad-TTS, LightGrad achieves 62.2\% reduction in parameters and 65.7\% reduction in latency, while preserving comparable speech quality in Chinese and English using 4 denoising steps.
%zywu-1006 已经解决
%可以不单独一句来说，而是作为上一句话的从句加以补充
%比如 while achieving (or preserving comparable speech quality on both Chinese Mandarin and English
%Experiment results on Chinese Mandarin and English show that LightGrad is able to synthesize speech of comparable quality.
% 之后打算把模型的代码开源

\section{Methodology}
\label{sec:metho}

% 1. 基于SDE的DDPM前向、后向过程描述
%   1.1 基于SDE的DDPM前向过程，描述和gradtts里面的方法一致
%   1.2 基于SDE的DDPM后向过程，描述和gradtts里面的方法一致
%   1.3 训练的loss
% 2. 快速采样方法 DPM-solver
% 3. 轻量化模型： 对depthwise separable convolution进行描述
% 4. 流式推理方法

% GradTTS: Number of encoder + duration predictor parameters: 7222033
%          Number of decoder parameters: 7634887
%          Total parameters: 14856920
% LightGrad: Number of encoder + duration predictor parameters: 3570129
%            Number of decoder parameters: 2041051
%            Total parameters: 5611180
Based on Grad-TTS, LightGrad is a non-autoregressive TTS model with an encoder, a duration predictor, and a lightweight U-Net decoder producing mel-spectrograms by gradually transforming noise sampled from prior distribution estimated by the encoder.
The lightweight U-Net employs depthwise separable convolutions to reduce parameters and computation.
Additionally, we leverage DPM-solver \cite{dpmsolver} to accelerate the sampling procedure in LightGrad.
Finally, streaming inference is implemented to reduce inference latency further.

\subsection{Background on diffusion probabilistic model}
\label{ssec1:method_dpm}
The basic idea of DPM is:
in forward process, we first convert data distribution to isotropic Gaussian by adding white noise gradually;
in backward process, a trained neural network restores data from Gaussian noise iteratively.
Follow \cite{sde,gradtts}, we define the forward process and backward process of LightGrad in terms of SDE.

\subsubsection{Forward process}
\label{sssec2:method_forward_process}
The forward process in LightGrad can be described as:
\begin{equation}
    dX_t=\frac{1}{2}(\mu-X_t)\beta_tdt+\sqrt{\beta_t}dW_t, t\in[0,T]
\end{equation}
where $\mu$ is the mean of Gaussian prior $\mathcal{N}(\mu,I)$, $\beta_t$ is a non-negative function referred as noise schedule and $W_t$ is the Brownian motion.
The forward process creates a stochastic process $\{X_t\}_{t=0}^T$. 
Assume $X_t\sim p(X_t)$, the forward process given above converts data distribution $p(X_0)$ into $p(X_T)\sim \mathcal{N}(\mu,I)$.
Given $X_0$, we can efficiently sample $X_t$ from:
\begin{equation}
\label{eq:sxt}
    p(X_t|X_0)=\mathcal{N}(\mu_t,\mathit{\Sigma_t})
\end{equation}
where 
\begin{equation}
    \mu_t=(I-e^{\frac{1}{2}\rho_t})\mu+e^{\frac{1}{2}\rho_t}X_0, \Sigma_t=I-e^{\rho_t}, \rho_t=-\int_0^t\beta_sds
\end{equation}

\subsubsection{Backward process}
\label{sssec3:method_backward_process}
To recover $X_0$ from $X_T$, we can solve backward SDE starting from $X_T$.
The backward process SDE is:
\begin{equation}
\label{solver1}
    dX_t=\Big(\frac{1}{2}(\mu-X_t)-\nabla \log p_t\Big)\beta_tdt+\sqrt{\beta_t}d\widetilde{W_t}
\end{equation}
where $\widetilde{W_t}$ is the reverse-time Brownian motion, $\nabla \log p_t$ is referred as \textit{score} of $p(X_t)$.
Additionally, \cite{sde} shows that (\ref{solver1}) shares the same marginal probability densities with an ordinary differential equation (ODE):
\begin{equation}
\label{solver2}
    dX_t=\frac{1}{2}\Big((\mu-X_t)-\nabla \log p_t\Big)\beta_tdt
\end{equation}
Thus, if we have a neural network trained to estimate $\nabla \log p_t$ for $t \in [0,T]$, we can sample $X_T$ from $p(X_T)$ and transform it to $X_0$ according to either (\ref{solver1}) or (\ref{solver2}).
\subsubsection{Training LightGrad}

To generate samples with DPM, a neural network $s_{\theta}$ is trained to estimate $\nabla \log p_t$ given $X_t$, $t$ and $\mu$.
When conditioned on $X_0$, we can sample $X_t$ directly using (\ref{eq:sxt}).
The score of $p(X_t|X_0)$ is:
\begin{equation}
    \nabla \log p(X_t|X_0)=-\frac{\epsilon_t}{\sqrt{\Sigma_t}}, \epsilon_t \sim \mathcal{N}(0,I)
\end{equation}
Thus, the corresponding diffusion loss function for training $s_{\theta}$ is:
\begin{equation}
    L_t=\mathbb{E}_{X_0,t}\Big[\mathbb{E}_{\epsilon_t}||\sqrt{\Sigma_t}s_\theta(X_t,\mu,t)+\epsilon_t||^2\Big]
\end{equation}

In addition to diffusion loss, similar to Grad-TTS, a negative log-likelihood encoder loss is applied on encoder outputs, and the duration predictor is trained using mean square error loss to estimate logarithmic duration.
% 这块MAS已经在introduction提过了全称，所以直接用缩写
The alignment between encoder outputs and target mel-spectrogram is estimated using MAS \cite{glowtts}.

\subsection{Fast sampling technique}
\label{ssec2:method_fast_sampling}
Sampling from DPMs can be regarded as solving corresponding backward SDE or ODE numerically, such as (\ref{solver1}) and (\ref{solver2}).
%As ODE can be solved with larger step size and efficient numerical ODE solvers, we use (\ref{solver2}) to sample from DPM.
In this paper, ODE is chosen for sampling from DPM.
(\ref{solver2}) have a semi-linear structure, which contains a linear function of data variable and a nonlinear function $s_\theta$.
General ODE solvers ignoring this semi-linear structure  cause discretization errors of both the linear and nonlinear term, preventing them from using larger \textit{step size} to enable fast and high-quality few-step sampling.
To improve sampling efficiency, we adopt DPM-Solver \cite{dpmsolver}, a fast dedicated solver for (\ref{solver2}).

Consider a neural network $s_{\theta}$ trained to estimate $\nabla \log p_t$.
Sampling process starts from $X_T\sim \mathcal{N}(\mu,I)$ and solves (\ref{solver2}) backward in time:
\begin{equation}
    dX_t=\frac{1}{2}\Big((\mu-X_t)-s_{\theta}(X_t,\mu,t)\Big) \beta_tdt
\end{equation}
Let $Y_t=X_t-\mu$, where $Y_T\sim \mathcal{N}(0,I)$, we have:
\begin{equation}
\label{solver3}
    dY_t=-\frac{1}{2}\beta_tY_t dt-\frac{1}{2}\beta_ts_{\theta}(Y_t+\mu,\mu,t) dt
\end{equation}
\cite{dpmsolver} reveals that (\ref{solver3}) has a semi-linear structure and consists of two parts: $-\frac{1}{2}\beta_t Y_t dt$ is a linear function of $Y_t$, and  $-\frac{1}{2}\beta_ts_{\theta}(Y_t+\mu,\mu,t)dt$ is a non-linear function of $Y_t$.
For $s \in (0,T)$ and $t \in [0,s]$, an exact solution of (\ref{solver3}) is:
\begin{equation}
    \label{solver4}
    Y_t=\frac{\alpha_t}{\alpha_s}Y_s+\alpha_t \int_{\lambda_s}^{\lambda_t}e^{-\lambda}\sqrt{\Sigma_{\tau_\lambda}}s_\theta(Y_{\tau_{\lambda}}+\mu,\mu,\tau_{\lambda})d\lambda
\end{equation}
where 
\begin{equation}
    \alpha_t=e^{\frac{1}{2}\rho_t}, \sigma_t=\sqrt{\Sigma_t}, \lambda_t=\lambda(t)=\log{\frac{\alpha_t}{\sigma_t}}, \tau_\lambda=\lambda^{-1}(\lambda)
\end{equation}
 Thus, given $Y_s$ at time $s$, the approximate solution for $Y_t$ is equivalent to approximating integral in (\ref{solver4}), which avoids the error of the linear term.
Substituting $s_\theta$ in the exponentially weighted integral with its Taylor expansion and approximating the first $(k-1)$-th total derivatives of $s_\theta$ can derive $k$-th-order ODE solver called DPM-Solver-$k$ \cite{dpmsolver}.
In this paper, we adopt DPM-Solver-1, which can be described as:
\begin{equation}
    Y_t=\frac{\alpha_t}{\alpha_s}Y_s+\sigma_t(e^{\lambda_t-\lambda_s}-1)\sqrt{\Sigma_s}s_\theta(Y_s+\mu,\mu,s)
\end{equation}
Detailed derivations of above equations can be found in \cite{dpmsolver}.

%\vspace{-0.7cm}
\begin{figure*}[!htb]
\centering
\vspace{-0.2cm}
\subfloat[Lightweight U-Net]{
\includegraphics[width=0.3\linewidth]{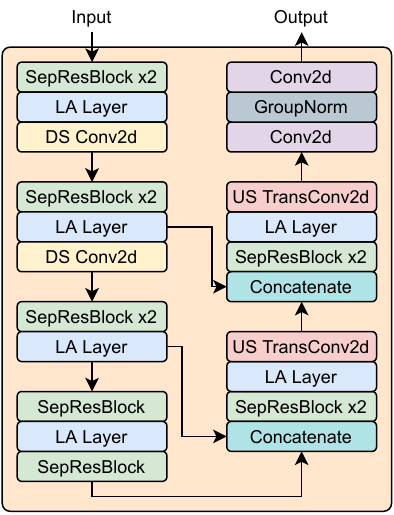}
\label{fig:lwunet}
}
\quad
\subfloat[Separable resnet block (SepResBlock)]{
\includegraphics[width=0.3\linewidth]{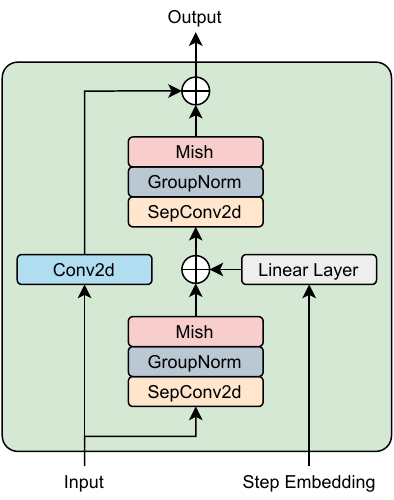}
\label{fig:sepresb}
}
\quad
\subfloat[Linear attention layer (LA layer)]{
\includegraphics[width=0.3\linewidth]{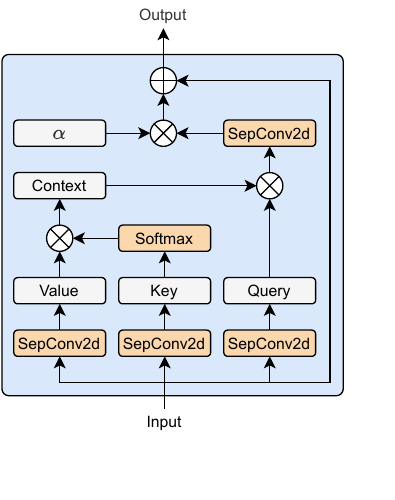}
\label{fig:lalayer}
}
\caption{The overall architecture for lightweight U-Net. For brevity, diffusion step embedding for SepResBlock in (a) is omitted.}
\label{fig:model_structure}
\vspace{-0.1cm}
\end{figure*}

\subsection{Lightweight U-Net}
\label{ssec3:metho_decoder}
%The decoder in LightGrad is a U-Net with residual connection and linear attention \cite{linearattention}, and we call this structure lightweight U-Net.
Generally speaking, the difference between our lightweight U-Net and the diffusion decoder in Grad-TTS is that regular convolutions are substituted with depthwise separable convolutions \cite{depthwise} to reduce model parameters and computation.
The proposed lightweight U-Net contains three downsampling blocks, one middle block, two upsampling blocks and one final convolution block, and its structure is shown in Fig.\ref{fig:model_structure}\subref{fig:lwunet}.
The main building blocks of downsampling and upsampling blocks are separable resnet block (SepResBlock) and linear attention layer (LA layer), whose structures are show in Fig.\ref{fig:model_structure}\subref{fig:sepresb} and Fig.\ref{fig:model_structure}\subref{fig:lalayer}, respectively.

%zywu-1006 已解决
%1）图1b的画法有问题：主干不清晰。不要为了追求规整而引起不清晰的问题。
%建议突出主干（2 SepConv2d, GroupNorm）
%2）另外，除了说做了什么设计之外，还要说明为什么要这么设计。比如：
%2.1）Step Embedding为什么要这么处理，没有说清楚？
%2.2）Step Embedding已经是包含mish act.的输出了，为什么还要再经过mish+linear进行处理？为啥可以加到1st groupnorm之上？
%2.3）为什么要加residual connection？有啥好处？
The SepResBlock has two 2D depthwise separable convolution (SepConv2d) followed by group nomalization and Mish activation function \cite{mish}.
Diffusion step is represented by sinusoidal position embedding, which is further transformed by two linear layers with Mish activation function to get step embedding.
%, step embedding is processed by a linear layer to change its shape and is added to the output of the first Mish function of SepResBlock.
%In SepResBlock, step embedding is processed by a linear layer to adjust its shape and is added to the output of the first Mish function of SepResBlock.
In each SepResBlock, an extra linear layer is used to adjust the shape of step embedding.
Later, step embedding is added to the output of the first Mish function to inject diffusion step information into SepResBlock.
%In SepResBlock, a linear layer is used to adjust
%To inject diffusion step information into SepResBlock, step embedding is passed to a l
%Additionally, the input of SepResBlock is added to the output of the second Mish function to create a residual connection, where a regular 2D convolution is applied to match the shape if necessary.
Additionally, a residual connection is created by adding the input of SepResBlock to the output of the second Mish function, where a 2D convolution is applied to match their shapes if necessary.

%zywu-1006 已解决
%LinearAttention的输出是啥？
%为什么Key Value先处理为Conext？不是Query, Key算权重，然后再和Value加权后得到输出？
The LA layer performs linear self-attention on its input \cite{linearattention}, whose complexity is a linear function of input sequence length.
The input is first passed to three SepConv2d to get query, key and value.
Then, key is processed by softmax and multiplied with value to get context.
Finally, context is multiplied with query to get attention output, which is processed by a SepConv2d to get the final output of attention layer.
Follow Grad-TTS, ReZero \cite{ReZero} is implemented in LA layer.

The downsampling block in lightweight U-Net contains two SepResBlocks and one LA layer followed by one regular convolution whose stride is 2 (DS Conv2d) performing downsampling.
%zywu-1006 这个不做出解释和修改，因为GradTTS就是这样设计的
%既然第3个block的downsampling操作去掉了，还能叫3个downsampling block么？
Similar to Grad-TTS, we removed downsampling operation in the last downsampling block.
The middle block is composed of two SepResBlocks with LA layer in between.
The upsampling block is similar to downsampling block except that transpose convolution with stride 2 (US TransConv2d) performs upsampling operation after LA layer.
%zywu-1006 已解决
%说明与原始U-net类似，up-有从down-过来的
Like original U-Net, each upsampling block also receives output of LA layer in downsampling block.
Finally, the output of the last upsampling layer is passed to two 2D convolutions with group normalization in between to produce the output of lightweight U-Net.

\subsection{Streaming inference}
\label{ssec4:method_streaming inference}
% 要解释一下何为streaming inference
To  decrease runtime memory usage and user perceived latency, streaming inference \cite{Ellinas20} is implemented in LightGrad.
First, decoder input is chopped into chunks at phoneme boundaries to cover several consecutive phonemes and the chunk lengths are limited to a predefined range.
To incorporate context information into decoder, last phoneme of the previous chunk and first phoneme of the following chunk are padded to the head and tail of the current chunk.
Then, the decoder generates mel-spectrogram for each padded chunk.
After this, mel-spectrogram frames corresponding to the padded phonemes are removed to reverse the changes to each chunk.
In this way, noise, introduced by chopping decoder input of one phoneme into different chunks, can be prevented in the generated speech.
Also, padded phonemes help generate more natural prosody through providing context information for the decoder.

\section{Experiment}
\label{sec:experiment}

\subsection{Datasets}
\label{ssec1:datasets}

\begin{table*}
\centering
\vspace{-0.2cm}
\caption{Model comparison. Latency (Lat), RTF and peak memory (Mem) are tested on a single CPU thread (Intel(R) Core(TM) i7-9700K).} 
\label{table:table1}
\begin{tabular}{@{}lccccccccc@{}}
\toprule
\multicolumn{1}{c}{\multirow{2}{*}{\textbf{Model}}} & \multirow{2}{*}{\textbf{NFE}} & \multicolumn{2}{c}{\textbf{MOS}}                        & \multicolumn{2}{c}{\textbf{MCD}}    & \multirow{2}{*}{\textbf{Lat(s)}} & \multirow{2}{*}{\textbf{RTF}} & \multirow{2}{*}{\textbf{Mem(MB)}} & \multirow{2}{*}{\textbf{Params(M)}} \\ \cmidrule(lr){3-6}
\multicolumn{1}{c}{}                                &                               & \textbf{Chinese}           & \textbf{English}           & \textbf{Chinese} & \textbf{English} &                                  &                               &                                   &                                     \\ \midrule
GT(reconstructed)                                   & -                             & 4.602($\pm$0.052)          & 4.705($\pm$0.048)          & -                & -                & -                                & -                             & -                                 & -                                   \\ \midrule
Grad-TTS                                             & 10                            & 3.940($\pm$0.067)          & \textbf{4.200($\pm$0.064)} & 4.754            & \textbf{4.583}   & 10.512                           & 2.358                         & 427.8                             & \multirow{2}{*}{14.85}              \\
                                                    & 4                             & 3.018($\pm$0.075)          & 2.920($\pm$0.067)          & 5.439            & 5.178            & 4.198                            & 0.942                         & 453.0                             &                                     \\ \midrule
LightGrad                                           & 10                            & \textbf{4.108($\pm$0.063)} & 4.158($\pm$0.062)          & \textbf{4.718}   & 4.639            & 9.308                            & 2.093                         & 413.4                             &                                     \\
                                                    & 4                             & 4.050($\pm$0.062)          & 3.940($\pm$0.056)          & 4.899            & 4.592            & 3.605                            & \textbf{0.81}                 & 462.5                             & \textbf{5.61}                       \\
\quad+streaming                                     & 4                             & 4.010($\pm$0.061)          & 3.925($\pm$0.055)          & 4.826            & 4.603            & \textbf{0.615}                   & -                             & \textbf{220.0}                    &                                     \\ \bottomrule
\end{tabular}
\vspace{-0.1cm}
\end{table*}

We evaluated LightGrad on both Chinese and English dataset.
For Chinese, we use a public speech dataset\footnote{https://www.data-baker.com/open\_source.html} containing 10,000 audio clips whose total length is nearly 12 hours.
Numbers of samples for training, validation and testing are 9,600, 200 and 200, respectively.
For English, we use LJSpeech \cite{ljspeech17} containing 13,100 English audio clips, whose total length is nearly 24 hours.
Numbers of samples for training, validation and testing are 12,229, 348 and 523, respectively. 
Audios from both two datasets are resampled to 22,050Hz, and are converted to 80-dimensional mel-spectrograms with the frame size 1,024 and the hop size 256.

\subsection{System setup}
We select Grad-TTS as our baseline and follow its original setup.
Our LightGrad consists of an encoder, a duration predictor and a lightweight U-Net decoder.
The architecture of the encoder and the duration predictor is the same as Grad-TTS, but the encoder's hidden size and number of convolution channels in encoder are 128 and 512 respectively.
LightGrad is trained for 1.7M iterations on a single GPU with batch size 16, and Adam is chosen as the optimizer with learning rate 0.0001.
$T$ for the forward process of LightGrad is set to 1, and we use the same noise schedule as Grad-TTS in LightGrad.
During inference, the temperature hyperparameter $\tau$ is set to 1.5 for both Grad-TTS and LightGrad, i.e. $X_T\sim \mathcal{N}(\mu,\frac{I}{1.5})$.
We use the number of function evaluations (NFE, a.k.a number of denoising steps) to represent the number of calls to the decoder when sampling from DPM, and we set NFE for Grad-TTS and LightGrad to 4 and 10.
When LightGrad performs streaming inference, the decoder generates 0.5 second mel-spectrogram chunk each time.
HiFi-GAN \cite{hifigan} is chosen as the vocoder converting mel-spectrograms to audios.

\subsection{Results and analysis}
\label{sssec_r:results}

To evaluate the speech quality of LightGrad, we conducted a subjective test to compare LightGrad with other systems, including speeches reconstructed from ground truth mel-spectrogram (GT(reconstructed)) and Grad-TTS.
Mean opinion score (MOS) is selected as the evaluation metric of synthesized speeches.
For each model we randomly select 20 samples from test set and present them to 20 subjects in random order.
Subjects were asked to rate the quality of synthesized speeches on a scale from 1 to 5 with 1 point increment in terms of naturalness, robustness and noise.
Audios that are more natural, have fewer pronunciation mistakes and less noise, are considered better.
We also conducted an objective evaluation using mel cepstral distortion (MCD).
Additionally, a runtime performance comparison between LightGrad and Grad-TTS is performed.
Average MCD, average latency, realtime factor (RTF) and runtime peak memory are calculated on the whole test set.

Experiment results are shown in Table \ref{table:table1}.
On Chinese dataset, compared with Grad-TTS, LightGrad achieves better MOS and comparable MCD given the same NFE.
On English dataset, LightGrad achieves comparable MOS and MCD.
As 4-denoising-step LightGrad can synthesize speeches of comparable quality as 10-denoising-step Grad-TTS, a 62.2\% reduction in parameters and a 65.7\% reduction in latency can be observed.
Also, the fast sampling technique adopted in LightGrad effectively reduces quality drops when using smaller NFE.
Finally, the experiment result demonstrates that streaming inference employed in LightGrad can reduce both runtime memory usage and latency without hurting much speech quality.

\subsection{Ablation study}

To show the effectiveness of designs in LightGrad, ablation studies are conducted, where 20 subjects were asked to rate the comparison mean opinion score (CMOS) for 20 samples from test set in terms of naturalness, robustness and noise on both Chinese and English dataset.
To validate the effectiveness of lightweight U-Net, we replace it with a different decoder composed of four feed-forward Transformer (FFT) block \cite{fastspeech2} having roughly the same number of parameters as lightweight U-Net.
We also substitute the fast sampling technique in LightGrad with the original sampling method in Grad-TTS to show benefits of the fast sampling technique.
NFE for all diffusion models in ablation studies is set to 4.

Results of ablation studies are shown in Table \ref{table:table2}.
We find that using FFT blocks as decoder will produce less natural audios that contain more noise and have more pronunciation problems.
It results in -0.202 and -0.208 CMOS on Chinese and English dataset, respectively.
When the fast sampling technique is removed, severe CMOS drop can be observed: -1.585 and -1.760 CMOS on Chinese and English dataset.
It shows that fast sampling technique is vital to reducing inference latency for DPMs while keeping the quality of generated speech.

\begin{table}
\centering
\caption{Ablation study}
\label{table:table2}
\begin{tabular}{@{}llcccl@{}}
\toprule
\multicolumn{1}{c}{\multirow{2}{*}{\textbf{Model}}} &  & \textbf{}        & \multicolumn{1}{l}{CMOS} & \textbf{}        &  \\ \cmidrule(l){2-6} 
\multicolumn{1}{c}{}                                &  & \textbf{Chinese} & \multicolumn{1}{l}{}     & \textbf{English} &  \\ \midrule
LightGrad                                           &  & 0.0000           &                          & 0.0000           &  \\ \midrule
\quad-diffusion                                     &  & -0.202           &                          & -0.208           &  \\
\quad-fast sampling                                 &  & -1.585           &                          & -1.760           &  \\ \bottomrule
\end{tabular}
\vspace{-0.1cm}
\end{table}
\section{Conclusion}
\label{sec:conclusion}

In this work, we proposed LightGrad, a lightweight DPM for TTS.
Equipped with a lightweight U-Net decoder, a fast sampling technique, and streaming inference, 
LightGrad achieves 62.2\% reduction in parameters, 65.7\% reduction in latency and can synthesize speech of comparable quality in both Chinese and English using 4 steps for denoising.

\textbf{Acknowledgement}: This work is supported by National Natural Science Foundation of China (62076144), Shenzhen Key Laboratory of next generation interactive media innovative technology (ZDSYS20210623092001004) and Shenzhen Science and Technology Program (WDZC20220816140515001).
\bibliographystyle{IEEEbib}
\clearpage
\bibliography{refs}

\begin{thebibliography}{10}

\bibitem{tacotron}
Yuxuan Wang, R.~J. Skerry{-}Ryan, Daisy Stanton, Yonghui Wu, Ron~J. Weiss,
  Navdeep Jaitly, Zongheng Yang, Ying Xiao, Zhifeng Chen, Samy Bengio, Quoc~V.
  Le, Yannis Agiomyrgiannakis, Rob Clark, and Rif~A. Saurous,
\newblock ``Tacotron: Towards end-to-end speech synthesis,''
\newblock in {\em Interspeech}, Francisco Lacerda, Ed., 2017, pp. 4006--4010.

\bibitem{transformertts}
Naihan Li, Shujie Liu, Yanqing Liu, Sheng Zhao, and Ming Liu,
\newblock ``Neural speech synthesis with transformer network,''
\newblock in {\em AAAI}, 2019, pp. 6706--6713.

\bibitem{fastspeech}
Yi~Ren, Yangjun Ruan, Xu~Tan, Tao Qin, Sheng Zhao, Zhou Zhao, and Tie{-}Yan
  Liu,
\newblock ``Fastspeech: Fast, robust and controllable text to speech,''
\newblock in {\em NeurIPS}, 2019.

\bibitem{fastspeech2}
Yi~Ren, Chenxu Hu, Xu~Tan, Tao Qin, Sheng Zhao, Zhou Zhao, and Tie{-}Yan Liu,
\newblock ``Fastspeech 2: Fast and high-quality end-to-end text to speech,''
\newblock in {\em ICLR}, 2021.

\bibitem{glowtts}
Jaehyeon Kim, Sungwon Kim, Jungil Kong, and Sungroh Yoon,
\newblock ``Glow-tts: {A} generative flow for text-to-speech via monotonic
  alignment search,''
\newblock in {\em NeurIPS}, 2020.

\bibitem{luo21}
Renqian Luo, Xu~Tan, Rui Wang, Tao Qin, Jinzhu Li, Sheng Zhao, Enhong Chen, and
  Tie{-}Yan Liu,
\newblock ``Lightspeech: Lightweight and fast text to speech with neural
  architecture search,''
\newblock in {\em ICASSP}. 2021, pp. 5699--5703, {IEEE}.

\bibitem{li21}
Song Li, Beibei Ouyang, Lin Li, and Qingyang Hong,
\newblock ``Light-tts: Lightweight multi-speaker multi-lingual
  text-to-speech,''
\newblock in {\em ICASSP}. 2021, pp. 8383--8387, {IEEE}.

\bibitem{Xiao22}
Yujia Xiao, Xi~Wang, Lei He, and Frank~K. Soong,
\newblock ``Improving fastspeech {TTS} with efficient self-attention and
  compact feed-forward network,''
\newblock in {\em ICASSP}. 2022, pp. 7472--7476, {IEEE}.

\bibitem{Kang21}
Minsu Kang, Jihyun Lee, Simin Kim, and Injung Kim,
\newblock ``Fast {DCTTS:} efficient deep convolutional text-to-speech,''
\newblock in {\em ICASSP}. 2021, pp. 7043--7047, {IEEE}.

\bibitem{Jan20}
Jan Vainer and Ondrej Dusek,
\newblock ``Speedyspeech: Efficient neural speech synthesis,''
\newblock in {\em Interspeech}, 2020, pp. 3575--3579.

\bibitem{wang21}
Disong Wang, Liqun Deng, Yang Zhang, Nianzu Zheng, Yu~Ting Yeung, Xiao Chen,
  Xunying Liu, and Helen Meng,
\newblock ``Fcl-taco2: Towards fast, controllable and lightweight
  text-to-speech synthesis,''
\newblock in {\em ICASSP}. 2021, pp. 5714--5718, {IEEE}.

\bibitem{devicetts}
Zhiying Huang, Hao Li, and Ming Lei,
\newblock ``Devicetts: {A} small-footprint, fast, stable network for on-device
  text-to-speech,''
\newblock {\em CoRR}, vol. abs/2010.15311, 2020.

\bibitem{Ren21}
Yi~Ren, Jinglin Liu, and Zhou Zhao,
\newblock ``Portaspeech: Portable and high-quality generative text-to-speech,''
\newblock in {\em NeurIPS}, 2021, pp. 13963--13974.

\bibitem{ddpm1}
Jascha Sohl{-}Dickstein, Eric~A. Weiss, Niru Maheswaranathan, and Surya
  Ganguli,
\newblock ``Deep unsupervised learning using nonequilibrium thermodynamics,''
\newblock in {\em ICML}, Francis~R. Bach and David~M. Blei, Eds., 2015,
  vol.~37, pp. 2256--2265.

\bibitem{ddpm2}
Jonathan Ho, Ajay Jain, and Pieter Abbeel,
\newblock ``Denoising diffusion probabilistic models,''
\newblock in {\em NeurIPS}, 2020.

\bibitem{ddpm3}
Prafulla Dhariwal and Alexander~Quinn Nichol,
\newblock ``Diffusion models beat gans on image synthesis,''
\newblock in {\em NeurIPS}, 2021.

\bibitem{diffwave}
Zhifeng Kong, Wei Ping, Jiaji Huang, Kexin Zhao, and Bryan Catanzaro,
\newblock ``Diffwave: {A} versatile diffusion model for audio synthesis,''
\newblock in {\em ICLR}, 2021.

\bibitem{wavegrad}
Nanxin Chen, Yu~Zhang, Heiga Zen, Ron~J. Weiss, Mohammad Norouzi, and William
  Chan,
\newblock ``Wavegrad: Estimating gradients for waveform generation,''
\newblock in {\em ICLR}, 2021.

\bibitem{gradtts}
Vadim Popov, Ivan Vovk, Vladimir Gogoryan, Tasnima Sadekova, and Mikhail~A.
  Kudinov,
\newblock ``Grad-tts: {A} diffusion probabilistic model for text-to-speech,''
\newblock in {\em ICML}, 2021, vol. 139, pp. 8599--8608.

\bibitem{difftts}
Myeonghun Jeong, Hyeongju Kim, Sung~Jun Cheon, Byoung~Jin Choi, and Nam~Soo
  Kim,
\newblock ``Diff-tts: {A} denoising diffusion model for text-to-speech,''
\newblock in {\em Interspeech}, 2021, pp. 3605--3609.

\bibitem{diffsinger}
Jinglin Liu, Chengxi Li, Yi~Ren, Feiyang Chen, and Zhou Zhao,
\newblock ``Diffsinger: Singing voice synthesis via shallow diffusion
  mechanism,''
\newblock in {\em AAAI}, 2022, pp. 11020--11028.

\bibitem{diffsvc}
Songxiang Liu, Yuewen Cao, Dan Su, and Helen Meng,
\newblock ``Diffsvc: {A} diffusion probabilistic model for singing voice
  conversion,''
\newblock in {\em ASRU}. 2021, pp. 741--748, {IEEE}.

\bibitem{gradvc}
Vadim Popov, Ivan Vovk, Vladimir Gogoryan, Tasnima Sadekova, Mikhail~Sergeevich
  Kudinov, and Jiansheng Wei,
\newblock ``Diffusion-based voice conversion with fast maximum likelihood
  sampling scheme,''
\newblock in {\em ICLR}, 2022.

\bibitem{sde}
Yang Song, Jascha Sohl{-}Dickstein, Diederik~P. Kingma, Abhishek Kumar, Stefano
  Ermon, and Ben Poole,
\newblock ``Score-based generative modeling through stochastic differential
  equations,''
\newblock in {\em ICLR}, 2021.

\bibitem{depthwise}
Fran{\c{c}}ois Chollet,
\newblock ``Xception: Deep learning with depthwise separable convolutions,''
\newblock in {\em {CVPR}}. 2017, pp. 1800--1807, {IEEE} Computer Society.

\bibitem{dpmsolver}
Cheng Lu, Yuhao Zhou, Fan Bao, Jianfei Chen, Chongxuan Li, and Jun Zhu,
\newblock ``Dpm-solver: {A} fast {ODE} solver for diffusion probabilistic model
  sampling in around 10 steps,''
\newblock {\em CoRR}, vol. abs/2206.00927, 2022.

\bibitem{Ellinas20}
Nikolaos Ellinas, Georgios Vamvoukakis, Konstantinos Markopoulos, Aimilios
  Chalamandaris, Georgia Maniati, Panos Kakoulidis, Spyros Raptis, June~Sig
  Sung, Hyoungmin Park, and Pirros Tsiakoulis,
\newblock ``High quality streaming speech synthesis with low,
  sentence-length-independent latency,''
\newblock in {\em Interspeech}. 2020, pp. 2022--2026, {ISCA}.

\bibitem{mish}
Diganta Misra,
\newblock ``Mish: {A} self regularized non-monotonic activation function,''
\newblock in {\em 31st British Machine Vision Conference 2020, {BMVC} 2020}.
  2020, {BMVA} Press.

\bibitem{linearattention}
Zhuoran Shen, Mingyuan Zhang, Haiyu Zhao, Shuai Yi, and Hongsheng Li,
\newblock ``Efficient attention: Attention with linear complexities,''
\newblock in {\em {IEEE} Winter Conference on Applications of Computer Vision,
  {WACV}}. 2021, pp. 3530--3538, {IEEE}.

\bibitem{ReZero}
Thomas Bachlechner, Bodhisattwa~Prasad Majumder, Huanru~Henry Mao, Gary
  Cottrell, and Julian~J. McAuley,
\newblock ``Rezero is all you need: fast convergence at large depth,''
\newblock in {\em Proceedings of the Thirty-Seventh Conference on Uncertainty
  in Artificial Intelligence, {UAI}}. 2021, vol. 161 of {\em Proceedings of
  Machine Learning Research}, pp. 1352--1361, {AUAI} Press.

\bibitem{ljspeech17}
Keith Ito and Linda Johnson,
\newblock ``The lj speech dataset,'' 2017.

\bibitem{hifigan}
Jungil Kong, Jaehyeon Kim, and Jaekyoung Bae,
\newblock ``Hifi-gan: Generative adversarial networks for efficient and high
  fidelity speech synthesis,''
\newblock in {\em NeurIPS}, 2020.

\end{thebibliography}

\end{document}